\documentclass[prl,showpacs,reprint]{revtex4-1}

\usepackage[centertags]{amsmath}
\usepackage{amsfonts}
\usepackage{amssymb}
\usepackage[sort&compress]{natbib}
\usepackage[hypertex]{hyperref}

\newcommand{\e}{\varepsilon}

\begin{document}

\title{Comment on ``Finite size corrections to the radiation reaction force in classical electrodynamics''}

\date{\today}

\author{P.O. Kazinski}

\email{kpo@phys.tsu.ru}

\affiliation{Physics Faculty, Tomsk State University, Tomsk, 634050 Russia\\
Institute of Monitoring of Climatic and Ecological Systems, SB RAS, Tomsk, 634055 Russia}


\maketitle

1. The authors of Letter \cite{FSCRRFCE} claim that they ``prove that leading order effect due to the finite radius $R$ of a spherically symmetric charge is order $R^2$ rather than order $R$ in any physical model, as widely claimed in the literature'' since ``symmetries prohibit linear corrections''. I shall show that this is an incorrect statement.

Indeed, according to the effective field theory approach exploited in \cite{FSCRRFCE}, we should augment the initial classical action of a point charge
by every possible local combination of fields, which does not spoil any symmetry of the initial model. In the case considered in \cite{FSCRRFCE}, these terms should have a mass dimension one or two. The authors assert that there is only one term Eq. (17) complying with these requirements. Its mass dimension is two, whence the main statement of the Letter follows. However, it is not difficult to find two more terms: (in the proper time parameterization)
\begin{equation}\label{rigid term}
    A=\int d\tau\ddot{x}_\mu \ddot{x}^\mu,\qquad B=\int d\tau\partial_\mu F^{\mu\nu} \dot{x}_\nu
\end{equation}
with dimensions one and two, respectively. The term $B$ is the first low energy correction to the form factor of a charged particle due to its finite size \cite{Foldy}. The term $A$ provides a counterexample to the main claim of Letter \cite{FSCRRFCE}.

The ``rigid'' relativistic terms like $A$ are well-known \cite{Pisar}. They appear naturally in studying higher dimensional generalizations of the Lorentz-Dirac (LD) equation \cite{higher}.  Also, if one smears the current of a point charge in a Lorentz-invariant manner \cite{CarBatUz}
\begin{equation}\label{regulariz}
    j^\mu(x)\rightarrow j^\mu_\e(x)=\Box_x \int d^4yG_\e(x-y)j^\mu(y),
\end{equation}
where $G_\e(x)=\theta(x^0)\delta(x^2-\e^2)/2\pi$, then the first correction to the LD equation is of order $\e$ and is obtained by a variation of the term $A$ (see Eq. (26), \cite{rrmm}). It is that term which is ``prohibited'' as the authors of the Letter claim.

2. In order to get rid of divergences, the authors of \cite{FSCRRFCE} use an improper regularization: ``one should regularize the divergences... by using... dimensional regularization which... sets all power-divergent integrals... to zero''. This assertion contradicts the standard renormalization procedure \cite{Collins}. We can take the power-like divergences to be equal to zero only if: i) we know that such terms can be canceled by appropriate counterterms added to the initial Lagrangian; ii) experiments or symmetries require that the coefficients at these terms vanish. Using their regularization scheme, the authors missed one possible divergent structure (the term $A$), which is not prohibited by symmetries and cannot be set to zero at will. In \cite{brane} it was proven that if one uses the regularization \eqref{regulariz}, which has a clear physical interpretation and does not spoil any symmetry, or some equivalent to it then all the arising divergences are Lagrangian and can be renormalized.

3. Another flaw is concerned with an ignorance of the relation between the regularization parameter $\Lambda$ and the characteristic size $R$ of a charged body. As it is given in \cite{FSCRRFCE}, the particle creates the electromagnetic field as a point object (the current is localized on a world-line). This field is substituted to the Lorentz-like force taken on the trajectory of the particle, i.e., taken on scales much lesser than $R$, where the particle can not be considered as a point. To make this procedure consistent, the regularization parameter characterizing the wave-length cutoff must be of the same order or even larger than $R$ so that the particle ``looks'' like a point object for the electromagnetic field. Then the expansion of the radiation reaction force in terms of $R$  rearranges (see Eq. (25) with $\Lambda\propto1/R$) and requires a more careful examination. The contributions of higher multipoles, which were discarded in \cite{FSCRRFCE}, can be greater than the terms (25), (26).

\end{document}